\documentclass[11pt]{article}
\usepackage{amsmath,amssymb,amstext,amsthm}
\usepackage{graphicx}
\usepackage{color}
 \numberwithin{equation}{section}
\begin{document}
\baselineskip 17pt
\title{\Large\bf Improving MF-DFA model with applications in precious metals market}

\author{{\large Zhongjun Wang$^{a}$,  Mengye Sun$^{a}$ and A. M. Elsawah$^{b,c, }$\footnote{Corresponding author. E-mail: a\_elsawah85@yahoo.com,~amelsawah@uic.edu.hk,~a.elsawah@zu.edu.eg}} \\
\\{\footnotesize $^a$ {\it Department of Statistics, Science School, Wuhan University of Technology, Wuhan 430070, China}}\\
{\footnotesize $^b$ {\it Beijing Normal University-Hong Kong Baptist University United International College, Zhuhai 519085, China}}\\
{\footnotesize $^c$  {\it Department of Mathematics, Faculty of Science, Zagazig University, Zagazig 44519, Egypt}}
}
\date{}
\maketitle
\begin{abstract}
With the aggravation of the global economic crisis and inflation, the precious metals with safe-haven
function have become more popular. An improved MF-DFA method is proposed to analyze price
fluctuations of the precious metals market. Based on the widely used multifractal detrended fluctuation
analysis method (MF-DFA), we compare these two methods and find that the Bi-OSW-MF-DFA
method possesses better efficiency. This article analyzes the degree of multifractality between spot gold
market and spot silver market as well as their risks. From the numerical results and figures, it is found
that two elements constitute the contributions in the formation of multifractality in time series and the
risk of the spot silver market is higher than that of the spot gold market. This attempt could lead to a
better understanding of complicated precious metals market.
\end{abstract}
\section{Introduction}
With the development of capital markets, the types of financial products have been increasing rapidly and precious metals trading have been becoming more active in a diversified portfolio. As the main investments in precious metals market, the market price fluctuations of gold and silver will inevitably cause widespread concern. Especially when the financial crisis comes, the precious metals with safe-haven function are favored by investors. However, a financial market is a complex system including various interior and exterior factors which make it difficult to understand and describe [1].

According to the existing researches, the price fluctuations of precious metals market is a process with non-linear characteristics which exhibits complex dynamics features. Note that, the time series of gold consumer price index of Chinese, Indian and Turkey market series are of multifractal nature [2]. The study of multifractality in the financial time series could reveal critical information about dynamics and complexity of the financial markets [3].

Mandelbrot, 1988 proposed a loose tentative definition for fractal, a fractal is a shape made of parts similar to the whole in some way [4]. In recent decades, the fractal theory has been developing quickly and has become a new technique for the study of some issues in financial market which is a very complex system. With the help of nonlinear system theory, fractal capital markets theory not only explains the market behavior of many complex phenomena, but also provides a way of quantitative analysis. The traditional rescaled range analysis method (R/S) was first proposed and introduced by Hurst who was an English hydrologist [5]. It is used to analyze the fractal characteristics and long-term persistence in time series, however this method is only applicable to monofractal. So [6] improved it and used R/S analysis with a $q-$th order height-height correlarion function to detect the multifractal nature of KOSPI.

It has been acknowledged that the rescaled range analysis (R/S) and the detrended fluctuation analysis (DFA) are the methods for monofractal.[7-9] .These two method are unable to capture complex dynamics in a time series or to characterize their scaling properties when the processes are governed by more than one scaling exponent [10] .Considering this, Kantelhardt et al. (2002) proposed Multifractal Detrended Fluctuation Analysis (MF-DFA) [11] which is an effective method to reliably research long-range correlations in nonstationary time series and at the same time it can avoid erroneous judgements of correlation. Currently, MF-DFA method has been widely used in the study of multifractal characteristics of financial market and has got some attractive results [12-15], it allows determining the correlation properties on large time scale [16]. Zunino et al.(2008) found that the multifractal degree for a broad range of stock markets is associated with the stage of their development [12]. Yuan and Zhuang used MF-DFA method to measure multifractailty of stock price index fluctuation [17]. Also, components of empirical multifractality in financial returns were investigated with MF-DFA method [18]. However, a possible drawback of the MF-DFA method is the occurrence of abrupt jumps in the detrended profile at the boundaries between the segments, since the fitting polynomials in neighboring segments are not related [16].

This paper is organized as follows. In Section 2, we present the brief overview of existing
research studies with pros and cons, and then improve the MF-DFA method. In Section 3,
we use this newly-built Binary Overlapped Sliding Window-Based MF-DFA (Bi-OSW-MF-
DFA) to detect the multifractal characteristics of spot gold and silver markets. Furthermore, a comparative study is done for the magnitude of the multifractality between these two markets and analyze their risk. Finally, we close through the conclusions and discussions in Section 4.

\section{Construction of a new method to improve MF-DFA method}
\subsection{Review of MF-DFA methodology}
Multifractal detrended fluctuation analysis has been widely used in many fields and it can reliably determine the multifractal scaling
behavior of time series [11] and consists of six steps as follows.
\begin{description}
  \item[Step 1.]  Suppose $x_k$ is a series of length $N$, calculate the 'profile' and get the series
  $$Y(i)=\sum_{k=1}^i(x_k-\bar{x}),~i=1,2,...,N,~\bar{x}=\frac{1}{N}\sum_{i=1}^Nx_i.$$
\item[Step 2.] Divide the profile $Y(i)$ into $N_s=\left[\frac{N}{s}\right]$­ non-overlapping segments of equal length
$s$ .Considered that the length $N$ of the series is often not a multiple of the considered time scale $s$
which a short part at the end of the profile may be left. In order not to disregard this part of the
series and to ensure data integrity, the same procedure is repeated starting from the opposite
end of the time series. Finally, we can obtained $2N_s$ segments. where $[.]$ means integer arithmetic.
\item[Step 3.] Calculate the local trend for each of the $2N_s$ segments by a least-square fit of the series.
Then determine the variance
$$F^2(v,s)=\frac{1}{s}\sum_{j=1}^s\left(Y((v-1)s+i)-y_v(i)\right)^2,~v=1,2,...,2N_s.$$
for each segment $v,~v= 1,...,N_s$ and
$$F^2(v,s)=\frac{1}{s}\sum_{j=1}^s\left(Y(N-(v-N_s)s+i)-y_v(i)\right)^2,~v=1,2,...,2N_s.$$
for $v= 2N_s+1,...,2N_s.$ Here, $y_v (i)$ is the fitting polynomial in box $v$. Because different order MF-DFAs differ in the capability
of eliminating trends in the series, linear (MF-DFA1), quadratic (MF-DFA2), cubic (MF-DFA3), or higher order polynomials
can be considered in the fitting procedure.
\item[Step 4.]  Generally, we are interested in how the generalized $q$ dependent fluctuation functions $F_q(s)$ depends on the time
scale $s$ for different values of $q$. The fluctuation function of $q$th order is computed by averaging over all the fluctuations of
the segments, defined as:
$$F_q(s)=\left(\frac{1}{2N_s}\sum_{v=1}^{2N_s}\left(F^2(s,v)\right)^{\frac{q}{2}}\right)^{\frac{1}{q}}.$$
The index variable $q$ can take any real non-zero value. For $q = 0,$ we calculate the fluctuation function as given below:
$$F_0(s)=\exp\left(\frac{1}{4N_s}\sum_{v=1}^{2N_s}\ln\left(F^2(s,v)\right)\right).$$
\item[Step 5.]  Finally, the scaling behavior of the fluctuation functions is determined by analyzing log plots of $F_q(s)$ versus  $s$ for each value of $q.$  If the series $x_k$ is long-range power-law correlated, $F_q(s)$ increases for large values of $s$ as a power-law, as follows:
$$F_q(s)\thicksim s^{h(q)}.$$
Here $h(q)$ is known as the generalized Hurst exponent and $h(2)$ is the well-known
Hurst exponent $H$. In general, if the time series is monofractal, $h(q)$ is independent of $q$ i.e. $\Delta h(q)=h(q_{\min})-h(q_{\max})= 0$
and if the time series is multifractal, $h(q)$ depends on $q.$ It means $\Delta h(q) > 0$ would signal the presence of multifractality in
the time series. Richer multifractality corresponds to higher variability of $h(q)$.  $\Delta h(q)$ is therefore referred to as the degree
of multifractality.

For positive values of $q$, $h(q)$ describes the behaviour of segments with large fluctuations while for negative values of $q,$
$h(q)$ describes the behaviour of segments with small fluctuations. In general, $h(q)$ is a decreasing monotonic function of $q$ for a stationary time series which means that relatively small fluctuations happen more often in the series than relatively
large ones. If $h(q) > 0.5,$ the fluctuations related to q are persistently auto-correlated, if $h(q) < 0.5,$ the fluctuations related
to $q$ are anti-persistently auto-correlated and if $h(q) = 0.5,$ the fluctuations related to $q$ display a random walk behaviour. See [3].
\item[Step 6.] Multifractality of a time series can also be characterized in terms of the multifractal scaling exponent $\tau(q)$ which is related
to the generalized Hurst exponent $h(q)$ through the relation
$$\tau(q)=qh(q)-1.$$
Here $\tau(q)$ represents the temporal structure of the time series as a function of moments $q$ and it reflects the scaling
dependence of small fluctuations for negative values of $q$ and large fluctuations for positive values of $q.$ If $\tau(q)$ is a linear
function of $q,$ the time series can be regarded as monofractal and if $\tau(q)$ has a nonlinear dependence on $q,$ then the series is
multifractal.

The complexity in a time series can be better captured through the singularity spectrum, $f(\alpha).\alpha$  and $f(\alpha)$ can be obtained
through a Legendre transform of $q$ and $\tau(q)$.
$$\alpha=\frac{d\tau(q)}{dq}=h(q)+qh'(q).$$
$$f(\alpha)=q\alpha(q)-\tau(q)=q^2h'(q)+1.$$
Both the exponent $\alpha$ and spectrum $f(\alpha)$ express the singularity of the price series. The width of singularity spectrum $f(\alpha)$  is always denoted as $\Delta\alpha,$ where $\Delta\alpha=\alpha_{\max}-\alpha_{\min}.$ The $\Delta\alpha$ is used to denote the uniform degree of the distribution of normalized prices for the whole fractal structure. The bigger of $\Delta\alpha,$ is the smaller even distribution probability measure is, and the more violent price fluctuations will usually be expected. And $\Delta\alpha=0$ corresponded to the situation of completely uniform distribution.  The plot of singular exponent $\alpha$   and singularity spectrum  $f(\alpha)$ can reflect the properties of probability distribution. If the series exhibits a simple monofractal scaling behavior, the value of singularity spectrum $f(\alpha)$  will be a constant; if the series exhibits a simple multifractal scaling behavior, the value of singularity spectrum $f(\alpha)$  will change dependently on the singular exponent $\alpha.$ The reader can refer to [13, 19, 20].
\end{description}
\subsection{Improving MF-DFA methodology}
In our paper, we use different detrending procedures to improve MF-DFA in two aspects and
named it as Bi-OSW-MF-DFA.
\\\textbf{Firstly.} We want to extract data just for one time, so we divide the profile $Y(i)$
into two sections equally, the length $n$ of these two sections both equal $\left\lfloor\frac{N}{2}\right\rfloor,$  where $\lfloor x\rfloor$ means the
largest integer not greater than $x,$ then we get two sub-series.\\
\textbf{ Secondly.} We divide both the two sub-series into $N^*_s=\left\lceil\frac{N-s}{s-l}\right\rceil,$ where $\lceil x\rceil$ means the smallest  integer not less than $x,$ overlapping segments of equal length $s,~l,~\left(0<l<\frac{s}{2}\right)$ is the overlapped length of neighboring segments. Since the length   of the sub-series is often not a multiple of the considered time scale, the last segment of the first profile will include a short part in the beginning of the second sub-series. In order to ensure data integrity, the same procedure is repeated starting from the opposite end of the second sub-series. Thereby, $2N^*_s$ segments are obtained altogether.
\subsection{The advantages of Bi-OSW-MF-DFA method}
In order to explain why our method performs better and illustrate specifically, we can easily see that the number of segments in
Bi-OSW-MF-DFA is half of that in MF-DFA which denotes as $2N^*_s=N_s.$  Therefore,
$$F_q(s)=\left(\frac{1}{2N^*_s}\sum_{v=1}^{2N^*_s}\left(F^2(s,v)\right)^{\frac{q}{2}}\right)^{\frac{1}{q}}=\left(\frac{1}{N_s}\sum_{v=1}^{N_s}\left(F^2(s,v)\right)^{\frac{q}{2}}\right)^{\frac{1}{q}}.$$
As well as, we use three figures
to show the different ways of dividing the intervals of profile $Y(i)$ between MF-DFA method and
Bi-OSW-MF-DFA method. From the following Figures 1, 2 and 3, it can be easily seen the different detrending procedures.

On one hand, to ensure data integrity, MF-DFA processes data both in sequential order and in
reversed order. However, there will produce some repeated information which may increase error
and increase the time of computation. Bi-OSW-MF-DFA method only extracts data for once which
saves the computer running time. On the other hand, [16] pointed out that the
unrelated fitting polynomials in neighboring segments may cause abrupt jumps in the detrended
profile at the boundaries between the segments. So a simple way to avoid these jumps would be
the calculation of $F_q (s)$ based on polynomial fits in overlapping windows.
In Section 3.2, we will use numerical simulations to prove our method: Bi-OSW-MF-DFA method
possesses better efficiency in detail.
\begin{center}
\begin{figure}[http]
\includegraphics[scale=0.55]{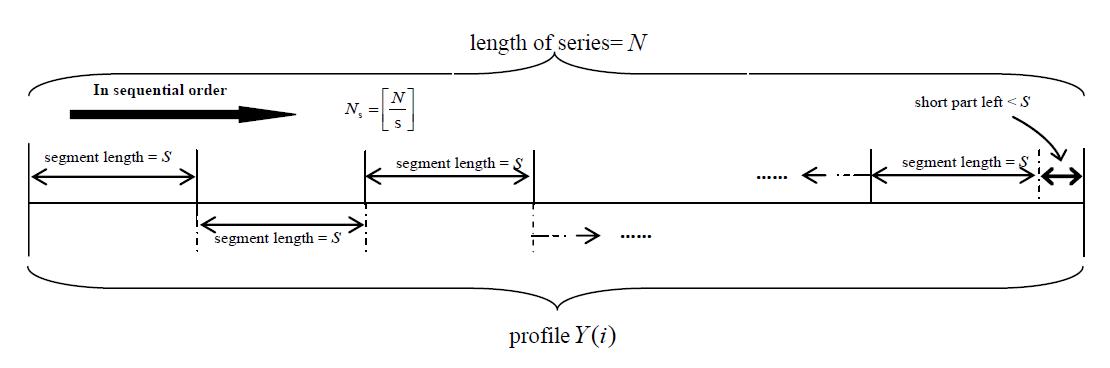}\\\vspace{-1.5cm}
\caption{ Way of dividing the intervals  $Y(i)$ in sequential order of MF-DFA method} \label{}
\end{figure}
\end{center}
\begin{center}
\begin{figure}[http]
\includegraphics[scale=0.55]{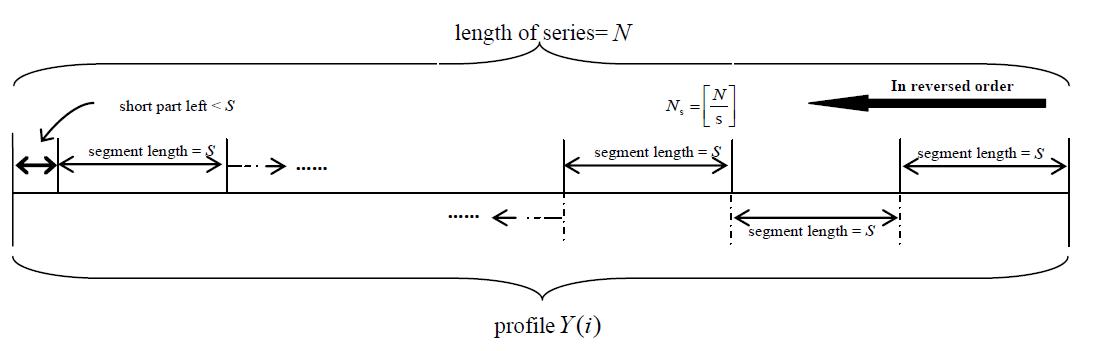}\\\vspace{-1.5cm}
\caption{ Way of dividing the intervals $Y(i)$ in reversed order of MF-DFA method} \label{}
\end{figure}
\end{center}
\begin{center}
\begin{figure}[http]
\includegraphics[scale=0.6]{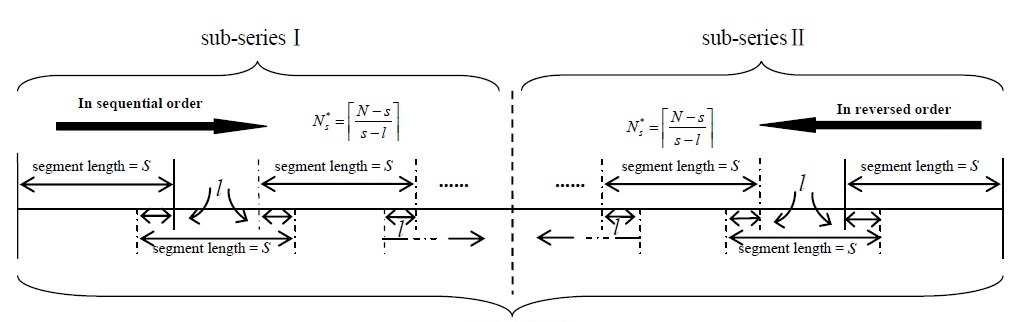}\\\vspace{-1.5cm}
\caption{ Way of dividing the intervals of profile $Y(i)$ of Bi-OSW-MF-DFA method} \label{}
\end{figure}
\end{center}
\section{Multifractal analysis and risk of gold and silver markets}
\subsection{Data description}
The main objective of this study is to investigate the characteristics of precious metals market,
using some multifractal measures to judge the market fluctuations. Daily closing price of spot gold
(from February 9, 2004 to March 23, 2015) and spot silver (from February 13, 2004 to March 23,
2015) are used as our samples. Eliminating weekends and holidays 2851 and 2880 different data
were obtained respectively. All these data have taken from the JIJIN silver and gold analysis
software. Figure 4 and Figure 5 show the daily closing price trends of spot gold and spot silver.

 As a first step, the unit root test is used to detect the stationarity of these two time series. Table 1 and
Table 2 show the results by using ADF test (defined below).\\
\textbf{ADF test:} In statistics and econometrics, an Augmented Dickey-Fuller test (ADF) is a test for a
unit root in a time series sample. Using unit root test could detect the stationarity of a time series [21].
\begin{table}[http]
\begin{center}
\caption{ADF test of gold daily closing price return series.}
\end{center}\vspace{-1cm}
 \scalebox{0.7}{\begin{tabular}[http]{|c|c|c|c|c|c|c|c|}
  \hline
  Variable&Test Type $(c,t,p)$&ADF Statistic&Threshold of $1\%$& Threshold of $5\%$&Threshold of $10\%$&Prob.& Conclusion\\\hline
  $Y$&$(c,0,0)$&$1.378$&$-2.568$&$-1.941$&$-1.617$&$0.958$&Nonstationary\\\hline
  $\Delta Y$&$(c,0,1)$&$-30.135$&$-2.568$&$-1.941$&$-1.617$&$0.000$&Stationary\\
  \hline
\end{tabular}}
\end{table}
\begin{table}[http]
\begin{center}
\caption{ADF test of silver daily closing price return series.}
\end{center}\vspace{-1cm}
 \scalebox{0.7}{\begin{tabular}[http]{|c|c|c|c|c|c|c|c|}
  \hline
  Variable&Test Type $(c,t,p)$&ADF Statistic&Threshold of $1\%$& Threshold of $5\%$&Threshold of $10\%$&Prob.& Conclusion\\\hline
  $Y$&$(c,0,0)$&$0.571 $&$-2.568 $&$-1.941 $&$-1.616 $&$0.839$&Nonstationary\\\hline
  $\Delta Y$&$(c,0,1)$&$-32.178 $&$-2.568$&$ -1.941 $&$-1.616$&$0.000$&Stationary\\
  \hline
\end{tabular}}
\end{table}
According to data in Tables 1 and 2 under the certain confidence level, statistic ADF $>$ critical value, the spot price of gold and silver are not stable and their first differences estimators are stationary. From the result of ADF test, we can learn that gold and silver spot price time series are non-stationary which also with a certain trend. Since logarithmic return rate can eliminate the dependence of price fluctuation on price level and return equals one order difference of logarithmic price series, we deal with the data in this way. Based on the original time
series, let $I_t$ be the closing price of time $t$ and $I_{t-1}$ be the closing price of last time. The
rate of return can be calculated as following formula
$$r_t=\ln I_t-\ln I_{t-1}.$$
Finally, we finally get 2850 and 2879  gold and silver daily return data respectively.

In order to obtain some further analysis of the data and get more accurate information, we use
SPSS to do descriptive statistics analysis of the return series as follows.
\begin{table}[http]
\begin{center}
\caption{Descriptive statistics of return series.}
\end{center}\vspace{-1cm}
\begin{center} {\begin{tabular}[http]{|c|c|c|c|c|c|c|}
  \hline
  Statistics&$N$& Mean& Minimum& Maximum &Skewness& Kurtosis\\\hline
  gold& 2879 &0.0003& -0.20 &0.13 &-1.074& 8.358\\\hline
  silver& 2850 &0.0004 &-0.11& 0.10 &-0.416& 6.706\\
  \hline
\end{tabular}}
\end{center}
\end{table}
\\From the Table 3, a positively large skew is seen. The skewness of spot gold and silver return time series are both
less than 0 and the peakedness are both greater than 3. The results also show that these two series
have a high degree of peakedness and fat tails and indicate that the return series of spot gold and
silver do not follow a normal distribution.

On the other hand, the following figures (see, Figures 6 and 7) describe the trends of spot gold and spot silver
daily rate of return.
\begin{center}
\begin{figure}[http]
\includegraphics[scale=0.9]{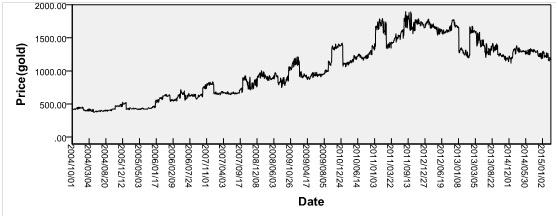}\\\vspace{-.5cm}
  \caption{Daily closing price history of spot gold market (2004-2015)} \label{}
\end{figure}
\end{center}
\begin{center}
\begin{figure}[http]
\includegraphics[scale=0.9]{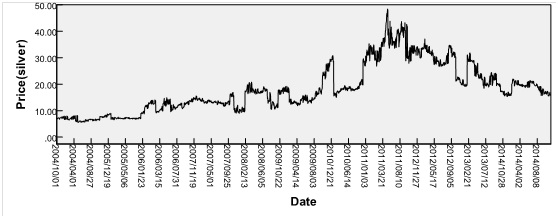}\\\vspace{-.5cm}
  \caption{Daily closing price history of spot silver market (2004-2015)} \label{}
\end{figure}
\end{center}
\begin{center}
\begin{figure}[http]
\includegraphics[scale=0.9]{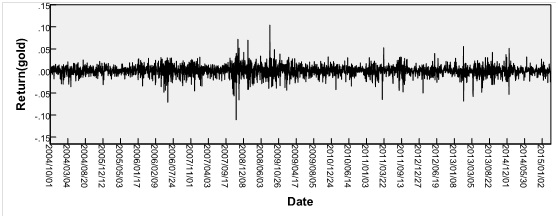}\\\vspace{-.5cm}
  \caption{Daily return series history of spot gold market (2004-2015)} \label{}
\end{figure}
\end{center}
\begin{center}
\begin{figure}[http]
\includegraphics[scale=0.9]{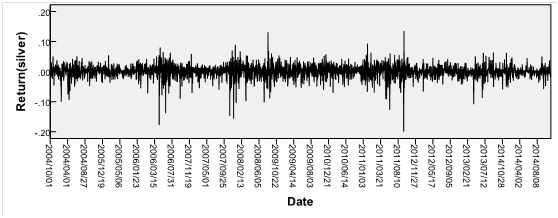}\\\vspace{-.5cm}
  \caption{Daily return series history of spot silver market (2004-2015)} \label{}
\end{figure}
\end{center}
\subsection{Comparison of MF-DFA and Bi-OSW-MF-DFA}
Compared with some mature markets such as the stock market, the future market and the bond market, precious metals market is still in its development stage. The precious metals market is a kind of complex non-liner dynamic system where the market information is asymmetrical. Presently the fluctuation of precious metals market seem quite confusing even to the regular traders, and it becomes almost impossible to predict its accurate rise and fall [2]. Although many natural phenomena seem rather complicated at the first sight, they do share some consistent and simple features [5].

On the other hand, Step 5 of MF-DFA has pointed out that if the series are long-range power-law correlated, the
fluctuation function $F_q (s)$ increases, for large value of $s$ , as a power-law,
$$F_q(s)\thicksim s^{h(q)}.$$
Note that, fluctuation function $F_q (s)$  can be defined only for $s\geq m+2$, where $m$ is the order of the
detrending polynomial. Moreover, $F_q (s)$ is statistically unstable for large scale $s (\geq N ).$ The generalized Hurst exponent $h(q)$ is extracted from a straight line fit to the log-log plot of $F_q (s)$ versus $s.$  See Figure 8 and 9.

Figures 8 and 9 give the results, we use moments with -20, -10, -2, + 2, +10 and +20 and the scale ranging from 100 to 500. Obviously, there exists some crossovers between different scaling regimes in the fluctuation functions . In this case a multitude of scaling exponents is required for a full description of the scaling behavior in the same range of time scales, and a multifractal analysis must be applied [11]. The following part we will use MF-DFA and Bi-OSW-MF-DFA to do multifractal analysis of spot gold and silver markets as well as the comparison between MF-DFA and Bi-OSW-MF-DFA.

To conform that the newly-built Bi-OSW-MF-DFA method is more robust than MF-DFA, we use
both methods to analyze the same gold and silver return series. In order to avoid errors caused by
differences of other parameters in multiple analysis, we select the same sub-interval length and
order where the time scale ranges from 10 to 50. Table 4 gives the Hurst exponent $h(q)$ calculated
from these two methods. For convenience, we denote $h(q)$ for gold return series and $h(q)$ for
silver return series in Table 4.

Because of $F_q(s)\thicksim s^{h(q)},$ it can be easily found that if pseudo fluctuations and some repeated
information are reducing, the errors will become less which leads to the decrease of fluctuation
functions $F_q(s)$ with different values of time scale $s$ . So the slope of log-log plot of the fluctuation
functions $F_q(s)$  which equals Hurst exponent $h(q)$ will become more concentrated and slighter.
Comparing the results from MF-DFA and Bi-OSW-MF-DFA, we find that to the gold return time
series: $\Delta {h_g^*}(q)=0.3860 < \Delta{h_g}(q)=0.5965$ to the silver return time series $\Delta {h_s^*}(q)=0.5515 <
\Delta {h_s}(q)=0.8501.$ These comparisons could manifest the (Bi-OSW-MF-DFA) method possesses
better robustness.
\begin{table}[http]
\begin{center}
\caption{Hurst exponent $h(q)$ calculated from MF-DFA and Bi-OSW-MF-DFA.}
\end{center}\vspace{-.5cm}
\begin{center} {\begin{tabular}[http]{|c|c|c|c|c|}
  \hline
  Order & \multicolumn{1}{c}~MF-DFA~&&\multicolumn{1}{c}Bi-OSW-MF-DFA&\\\hline
  q&$h_g(q)$ &$h_s(q)$  &${h_g^*}(q)$ &${h_s^*}(q)$ \\\hline
-20 &0.8704 &1.0400& 0.7217 &0.7849\\\hline
-16& 0.8569& 1.0237 &0.7129& 0.7757\\\hline
-12& 0.8331& 0.9942& 0.7005 &0.7619\\\hline
-8 &0.7832 &0.9287 &0.6807 &0.7398\\\hline
-4& 0.6887 &0.7729 &0.6387 &0.6856\\\hline
0 &0.5908& 0.6207 &0.5546 &0.5806\\\hline
2 &0.5316 &0.5480& 0.5129 &0.516\\\hline
4& 0.4634 &0.4326 &0.4719& 0.4302\\\hline
8& 0.3629& 0.2833 &0.4034& 0.3221\\\hline
12& 0.3139& 0.2294 &0.3674& 0.2748\\\hline
16& 0.2885& 0.2042 &0.3477 &0.2492\\\hline
20& 0.2739& 0.1899 &0.3357& 0.2334\\\hline
$\Delta h$ &0.5965 &0.8501 &0.3860 &0.5515\\\hline
\end{tabular}}
\end{center}
\end{table}\\
Figures 10 and 11 provide the relation between Hurst exponent $h(q)$
and order $q.$ By comparing the results of Figures 10 and  11, we can suggest that, with
Bi-OSW-MF-DFA method the Hurst exponent is more concentrated and the oscillation of Hurst
exponent is slighter. From the Figures 10 and 11, when assign the order $q$ from -20 to 20, the Hurst exponent $h(q)$ is
changing dependently. It indicates that, both in gold and silver markets, local trends of these two
series are not uniform, they do not exhibit a simple monofractal scaling behavior but with
multifractal characteristics. Meanwhile, the volatility of the generalized Hurst exponent can
measure the degree of multifractality in a financial market the more obvious, the greater its risk is.
Compared with silver, the range of gold return series generalized Hurst exponent $\Delta h_g^*(q)=0.3860$ is
less than the return series of silver $\Delta h_s^*(q)= 0.5515,$ so we can get, the
multifractality of silver return series is stronger and the risk of silver market is higher than gold
market.
\begin{center}
\begin{figure}[http]
\includegraphics[scale=0.45]{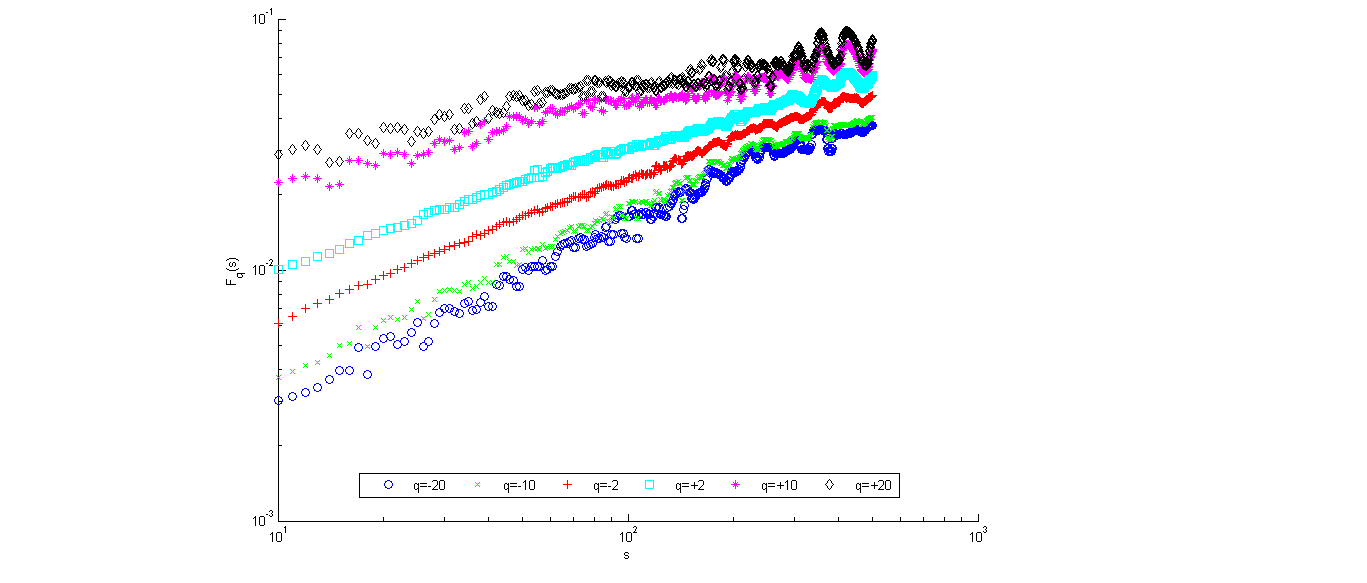}\\\vspace{-1cm}
  \caption{ $F_q (s)$ obtained from Bi-OSW-MF-DFA for gold return series for some
$q$-orders.} \label{}
\end{figure}
\end{center}
\begin{center}
\begin{figure}[http]
\includegraphics[scale=0.45]{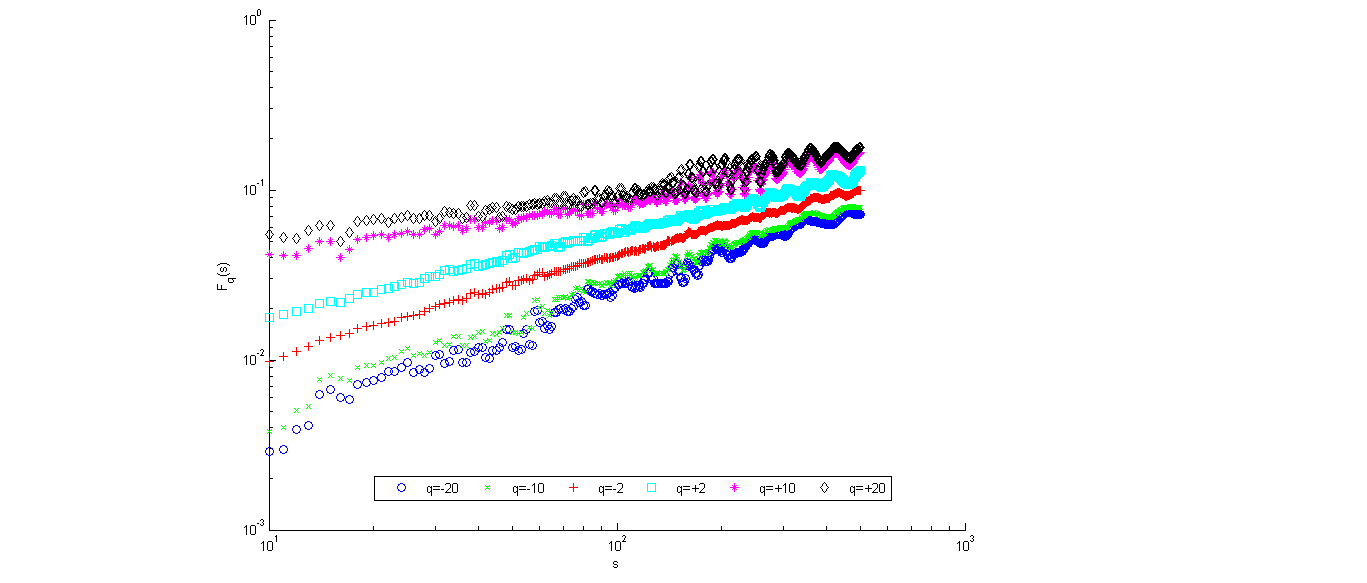}\\\vspace{-1cm}
  \caption{$F_q (s)$ obtained from Bi-OSW-MF-DFA for silver return series for some
$q$-orders.} \label{}
\end{figure}
\end{center}
\begin{center}
\begin{figure}[http]
\includegraphics[scale=0.45]{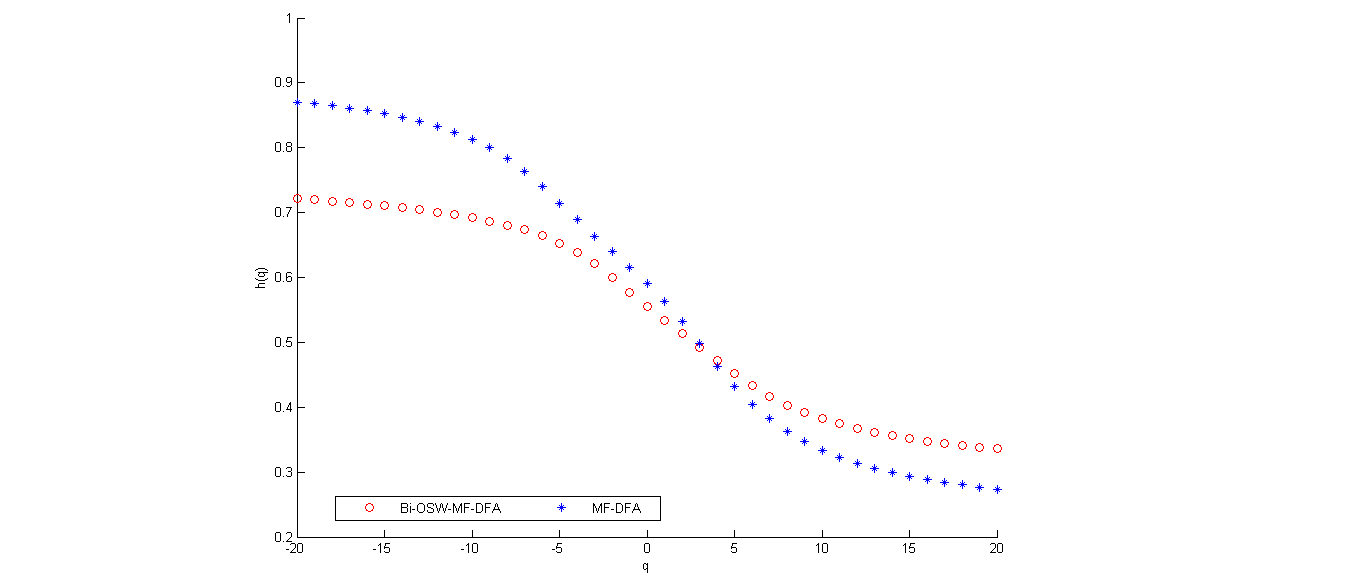}\\\vspace{-1cm}
  \caption{ $h(q)$ versus $q$ of the gold return series based on the two methods.} \label{}
\end{figure}
\end{center}
\begin{center}
\begin{figure}[http]
\includegraphics[scale=0.45]{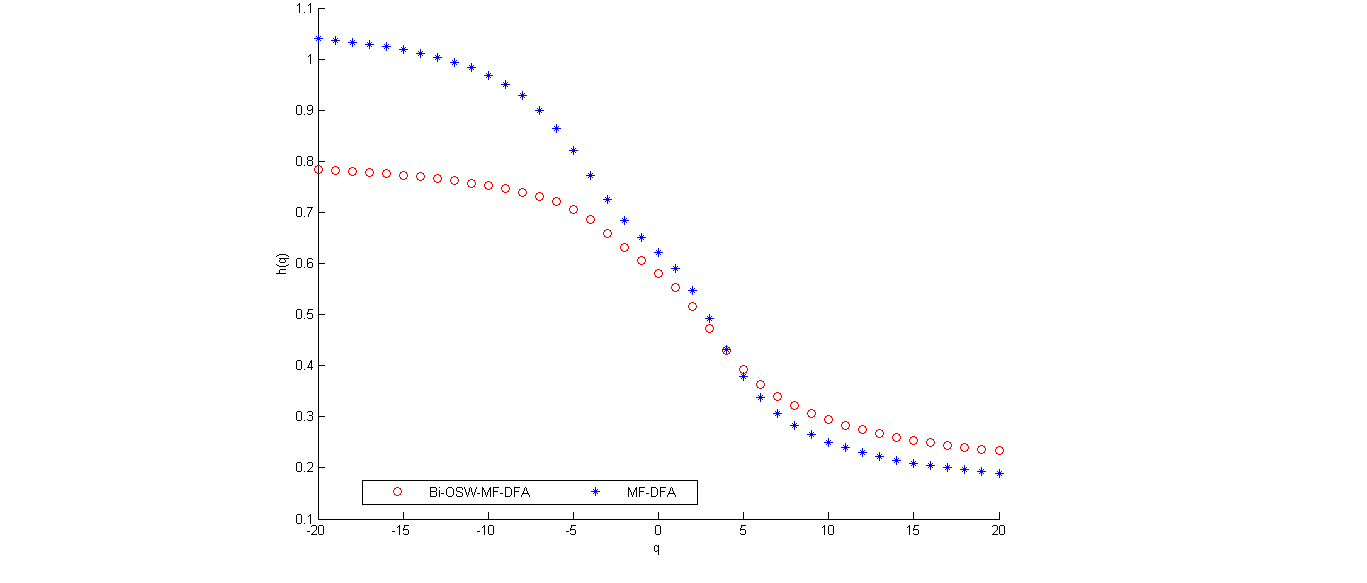}\\\vspace{-1cm}
  \caption{$h(q)$ versus  $q$ of the silver return series based on the two methods.} \label{}
\end{figure}
\end{center}
\subsection{Factors of multifractal structure in precious metals market}
According to some existing literatures, we have found that there are two different types of
multifractality in time series. (i) Multifractality due to a broad probability density function for the
values of the time series. (ii) Multifractality due to different long-range correlations of small and
large fluctuations [10]. In order to get further investigation of precious metals market
multifractality, we will respectively process in shuffling procedure and phase randomization
procedure.\\
\textbf{Shuffling procedure consists of three steps [14]:}
\begin{description}
  \item[Step1:] Generate pairs $(m,n)$ of random integer numbers (with $m,~n\leq N$) where $N$ is the total length of the time series to be shuffled.
  \item[Step2:] Interchange entries $m$ and $n.$
  \item[Step3:] Repeat steps 1 and 2 for $20N$ steps. (This step ensures that ordering of entries in the time series is fully shuffled.)
\end{description}
\textbf{Phase randomization procedure consists of three steps:}
 \begin{description}
  \item[Step1:] Process the original series with discrete Fourier transform.
  \item[Step2:] Rotate the phase with one phase angle.
  \item[Step3:] Then process the data with inverse Fourier transform.
\end{description}
Compared with the original series, the shuffled time series destroys correlations but can retain its
volatility, while the surrogate time series weakens the non-Gaussian distribution of time series.
Using the more efficient Bi-OSW-MF-DFA method to analyze the original series, the shuffled
series and the surrogate series of spot gold and silver markets. The results are as in Figures 14 and
15 and Table 5.

It is can be seen from Figures 14 and 15 that the Hurst exponent of gold and silver shuffled
series are both decreasing sharply. The changes indicate that market price fluctuations dominantly
affect the multifractality of gold and silver market. Comparing the two markets: the Hurst
exponent of gold shuffled series is farther from 0.5. After destroying the correlations, gold market
shows a stronger anti-persistence and this result proves that gold return series has a higher degree
of correlation.

Additionally, in view of Table 5, we have the $\Delta h = 0.1267$ of gold surrogate series is significantly less than the
silver surrogate series $\Delta h = 0.1604$ indicating that the fat-tailed probability distribution of the
gold market fluctuations leaves a deeper influence on its multifractality than the silver market.
\begin{center}
\begin{figure}[http]
\includegraphics[scale=0.43]{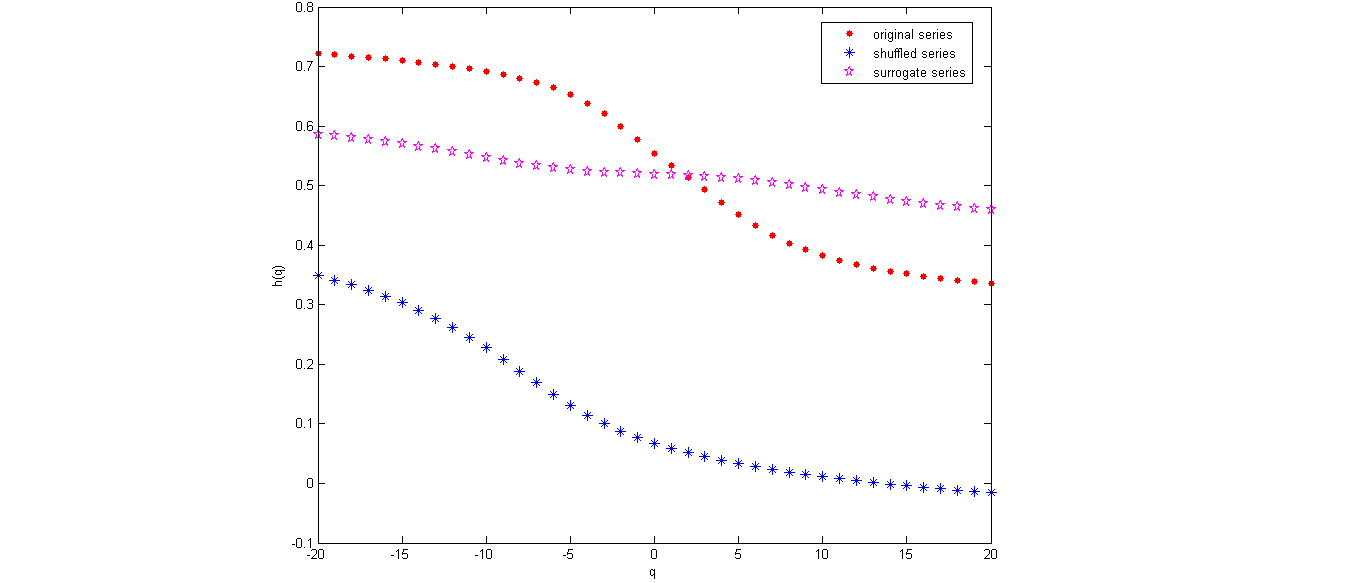}\\\vspace{-1cm}
  \caption{ $h(q)$ versus $q$ of original, shuffled and surrogate series in gold market based on Bi-OSW-MF-DFA method.} \label{}
\end{figure}
\end{center}
\begin{center}
\begin{figure}[http]
\includegraphics[scale=0.43]{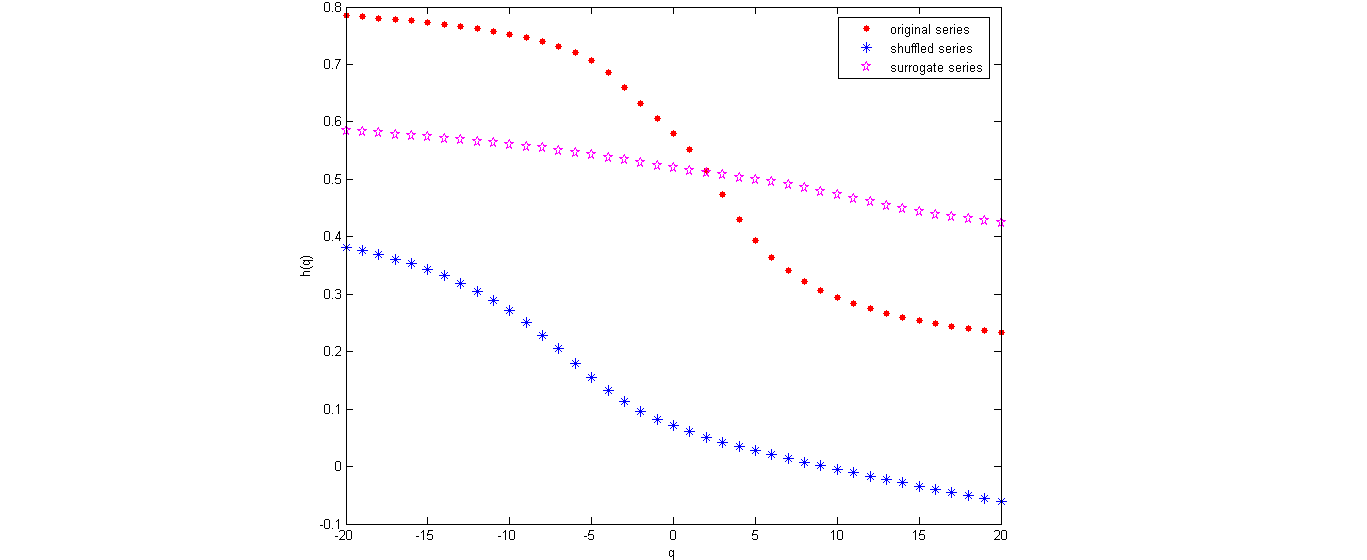}\\\vspace{-1cm}
  \caption{$h(q)$ versus $q$ of original, shuffled and surrogate series in silver market based on Bi-OSW-MF-DFA method.} \label{}
\end{figure}
\end{center}
\begin{table}[http]
\begin{center}
\caption{$h(q)$ of the original series, the shuffled series and the surrogate series.}
\end{center}\vspace{-1cm}
\begin{center}
\scalebox{.7}{\begin{tabular}[http]{|c|c|c|c|c|c|c|}
  \hline
  Order &\multicolumn{2}{c} { ~~~~~~~~~~~~~~~~~$h(q)$ of gold return series}&& \multicolumn{2}{c} {~~~~~~~~~~~~~~~$h(q)$ of silver return series}&\\\hline
  q&original series&shuffled series &surrogate series &original series &shuffled series &surrogate series\\\hline
-20& 0.7217& 0.3489& 0.5862& 0.7849 &0.3818 &0.5848\\\hline
-16& 0.7129 &0.3144 &0.5743 &0.7757 &0.3526 &0.5766\\\hline
-12& 0.7005 &0.2619 &0.5577 &0.7619 &0.3052 &0.5667\\\hline
-8 &0.6807 &0.1887 &0.5379 &0.7398 &0.2289 &0.5546\\\hline
-4 &0.6387 &0.1147 &0.5243& 0.6856 &0.1322 &0.5385\\\hline
0 &0.5546 &0.0672 &0.5195 &0.5806 &0.0708 &0.5205\\\hline
2 &0.5129 &0.0514 &0.5176 &0.5160& 0.0512 &0.5121\\\hline
4 &0.4719 &0.0386 &0.5142 &0.4302 &0.0349 &0.5040\\\hline
8 &0.4034 &0.0190 &0.5015 &0.3221 &0.0076 &0.4853\\\hline
12& 0.3674 &0.0044 &0.4852 &0.2748 &-0.0167& 0.4611\\\hline
16& 0.3477 &-0.0068& 0.4706& 0.2492 &-0.0397& 0.4395\\\hline
20& 0.3357 &-0.0156& 0.4595 &0.2334 &-0.0605 &0.4244\\\hline
$\Delta h$ &0.3860 &0.3645 &0.1267 &0.5515 &0.4423 &0.1604\\\hline
\end{tabular}}
\end{center}
\end{table}
\subsection{Scaling analysis of precious metals market}
With the relation between $h(q)$ and
$\tau(q)$, we can get the figure that $\tau(q)$ is changing with different $q.$ Since
$h(q)$ is a constant in monofractal analysis, the $\tau(q)\sim q $ chart is linear. And $h(q)$ will
change with different $q$ in multifractal analysis, so the curve is nonlinear.

We provide the $\tau(q)\sim q $  chart of these three different return series (see, Figures 16 and 17). A strong nonlinearity between $\tau(q)$ and $q$ can be easily seen from Figures 16 and 17. At the same time, both in gold and silver return series, the nonlinearity is more
obvious in the shuffled series which makes a more significant contribution to multiple scale.
Because of a stronger nonlinearity existing in silver return series, the price of silver market is
more likely to vibrate suddenly and unexpectedly.
\begin{center}
\begin{figure}[http]
\includegraphics[scale=0.45]{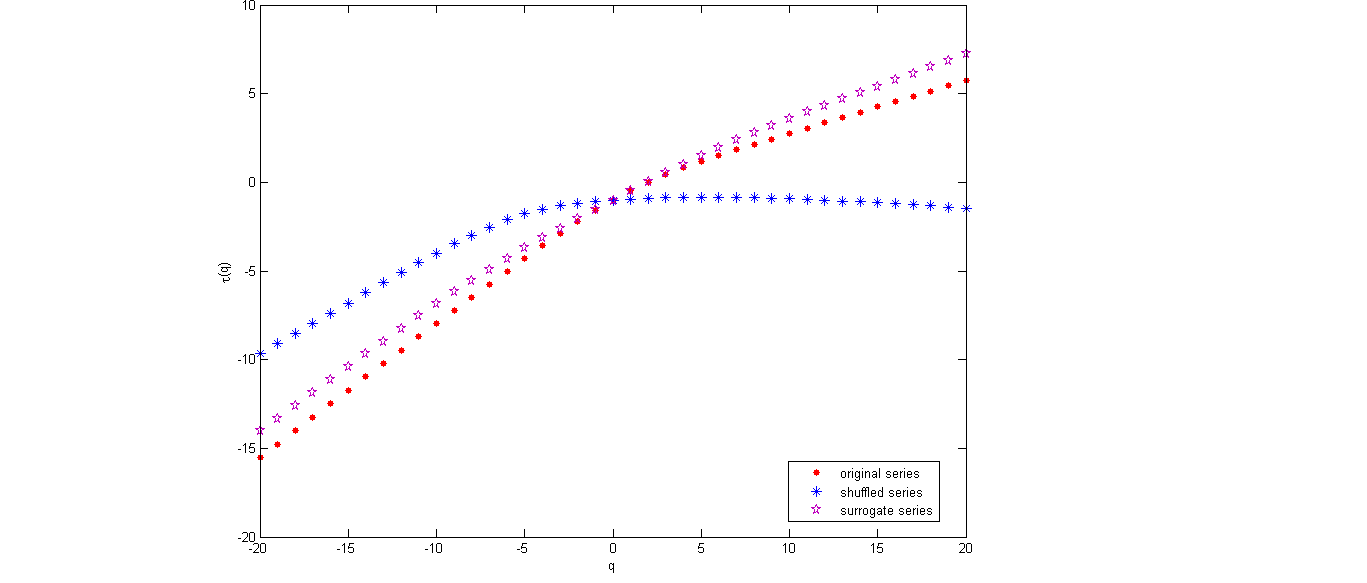}\\\vspace{-1cm}
  \caption{$\tau(q)$ versus the order $q$ of original, shuffled and surrogate series in gold market based on Bi-OSW-MF-DFA method.} \label{}
\end{figure}
\end{center}
\begin{center}
\begin{figure}[http]
\includegraphics[scale=0.45]{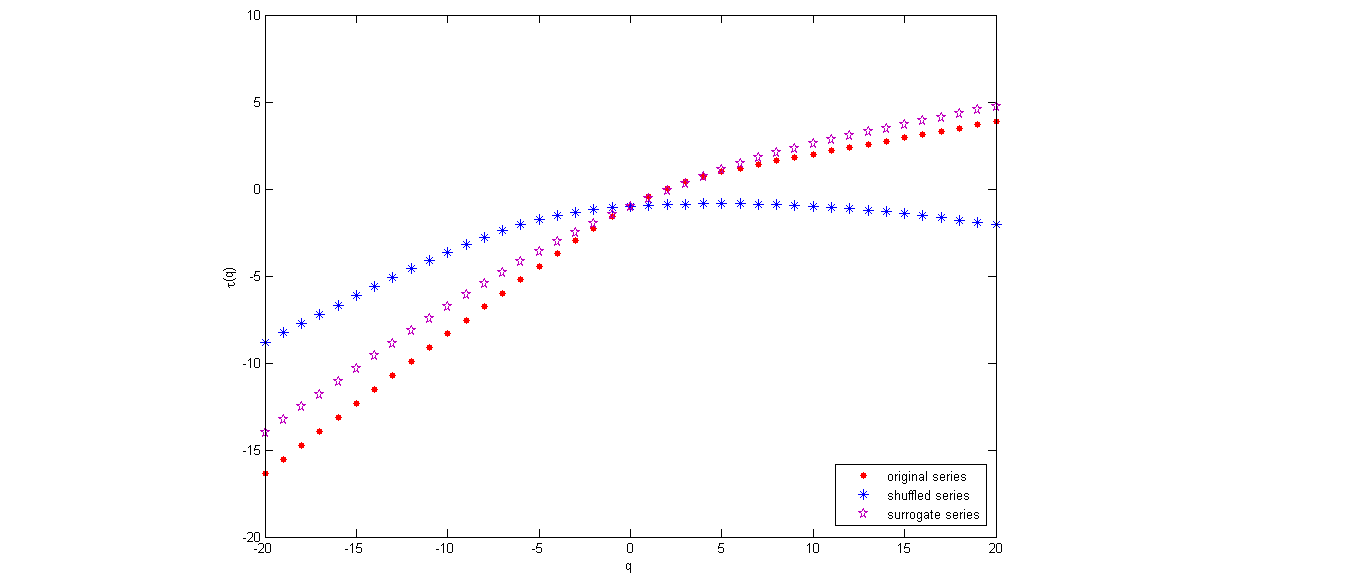}\\\vspace{-1cm}
  \caption{$\tau(q)$ versus the order $q$ of  original, shuffled and surrogate series in silver market based on Bi-OSW-MF-DFA method.} \label{}
\end{figure}
\end{center}
\subsection{Multifractal Spectrum Analysis of precious metals market risk}
We can obtain the plot of singular exponent $\alpha$ and multifractal spectrum $f(\alpha)$ with these
two formula $\alpha=h(q)+qh'(q)$ and $f(\alpha)=q^2h'(q)+1.$\\ The width of multifractal spectrum could reflect the risk of financial market. We can see from the Figures 18 and 19 and Table 6 that the width of the original silver return series multifractal spectrum $ \Delta \alpha$ equals 0.6160 is greater than it of the gold market which equals 0.4625. In silver market, there is a stronger price fluctuation which manifests the higher risk. Although the risk is more significant, but it also represents that we can get more profits. Generally speaking, silver market is suitable for speculators to get into. In both of gold and silver markets, there exist obvious differences of the multifractal spectrum width between the shuffled series and the surrogate series which confirms once again the multifractality in precious metals market is determined by two factors. In addition, the width of multifractal spectrum becomes narrower after phase randomization procedure indicates that extreme non-Gaussian events also have affected the multifractal properties in time series.
\begin{table}[http]
\begin{center}
\caption{Multifractality degree of three different series in both gold and silver markets.}
\end{center}\vspace{-1cm}
\begin{center} {\begin{tabular}[http]{|c|c|c|c|c|c|}
  \hline
  spot gold&$\Delta h$&$\Delta \alpha$&spot silver&$\Delta h$&$\Delta \alpha$\\\hline
  original series& 0.3860& 0.4625& original series &0.5515 &0.6160\\\hline
shuffled series &0.3645& 0.3629 &shuffled series &0.4423 &0.3947\\\hline
surrogate series &0.1267 &0.3178& surrogate series& 0.1604& 0.3835\\\hline
  \hline
\end{tabular}}
\end{center}
\end{table}
\begin{center}
\begin{figure}[http]
\includegraphics[scale=0.5]{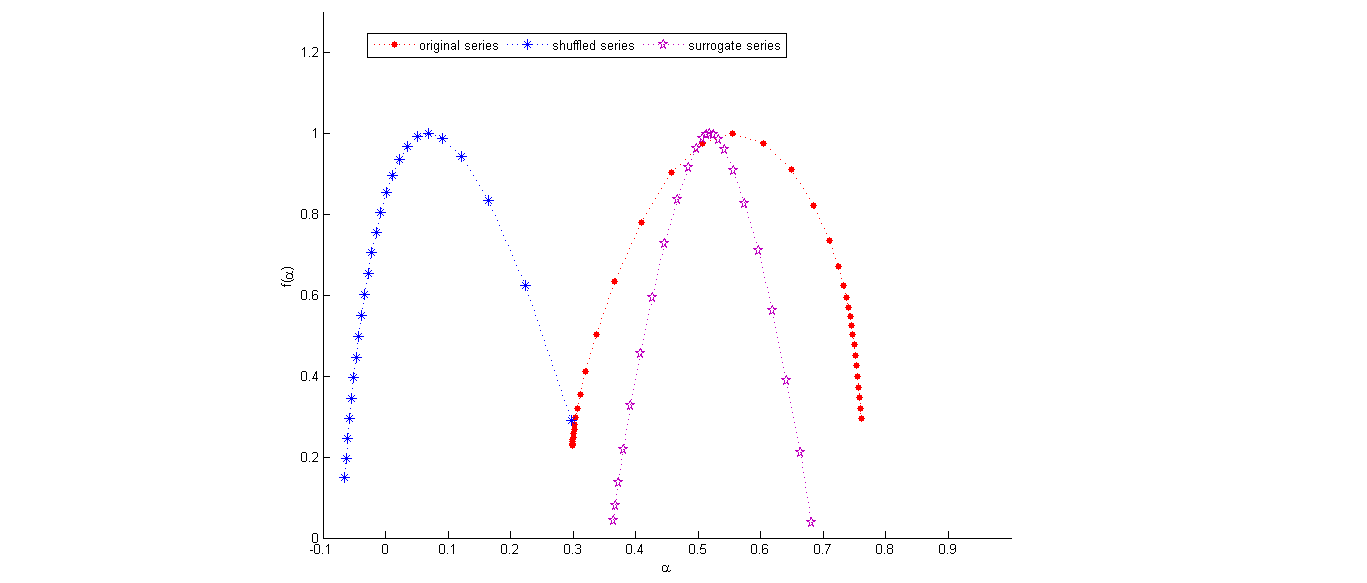}\\\vspace{-1cm}
  \caption{Multifractal spectrum $f(\alpha)$ of  original, shuffled and surrogate series in gold
market based on Bi-OSW-MF-DFA method.} \label{}
\end{figure}
\end{center}
\begin{center}
\begin{figure}[http]
\includegraphics[scale=0.5]{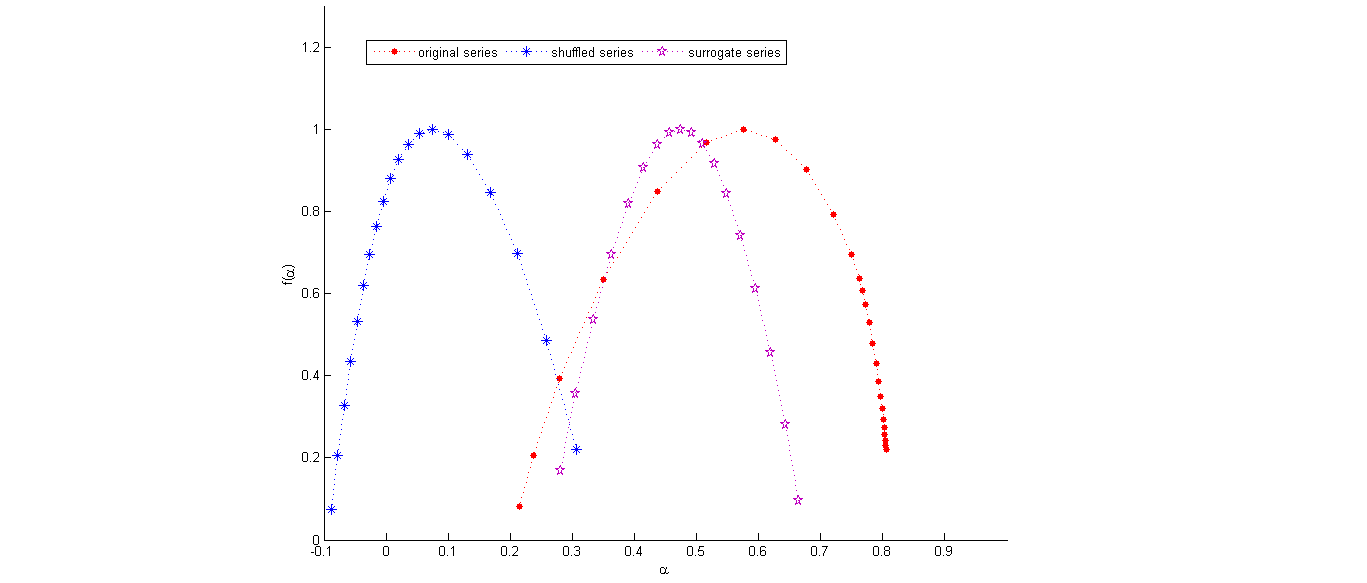}\\\vspace{-1cm}
  \caption{Multifractal spectrum $f(\alpha)$ of original, shuffled and surrogate series in silver
market based on Bi-OSW-MF-DFA method.} \label{}
\end{figure}
\end{center}
\section{Conclusions and discussions }
Based on fractal theory, this paper improves the MF-DFA method which is widely used in many fields and analyzes the precious metals market. We select the spot gold and silver market to do the research and obtain the following conclusions.
\begin{enumerate}
  \item We improve Multifractal Detrended Fluctuation Analysis (MF-DFA method) and obtain a better model--Bi-OSW-MF-DFA method. Binary Overlapped Sliding Window-based Multifractal Detrended Fluctuation Analysis possesses stronger robustness. This model
not only reduces the extraction of repeated information which decreases error, but also speeds up
operation in simulation.
  \item There are some crossovers both in the fluctuation function $F_q(s)$ of spot gold market and spot
silver market. Precious metals market does not exhibit a simple monofractal scaling behavior but
with significant multifractal properties. The statistics suggest that the multifractality in gold
market is less obvious than silver market which of more profitts and higher risk for investors.
  \item After shuffling procedure and phase randomization procedure, we can find that the multifractal
properties of the precious metals market is jointly caused by small and large fluctuations and volatility of the
different long-range correlation and volatility fat tail probability distribution, and the volatility of
time series is the main factor.
  \item Multifractal spectrum analysis of gold and silver markets shows that the multifractal spectrum width of silver series is wider than gold series, so the potential risk is higher in silver market. The wider the multifractal spectrum width is, the more significant the oscillation of market and the higher the rate of return; on the contrary, the less significant the oscillation of market is, the less risk exists. All the evidences suggest that there is a higher degree of security in gold market which makes it more suitable for investors to hedge.
\end{enumerate}
In real transaction, the smaller volume and the more speculative tendencies of investors lead to more serious price oscillations in silver market, and we have got the same conclusion by analyzing empirical data. Moreover, dispersion, high costs and other factors also make silver market price fluctuate strongly. But comparing with the stock market, precious metals market has a shorter history, is still an immature developing financial market which needs to be researched more deeply. Thus, more valuable and comprehensive information could be provided for investors to investigate the precious metals market.
\section*{Acknowledgments}
This work   was partially supported by the UIC Grants (Nos. R201810,  R201912 and R202010), the Zhuhai Premier Discipline Grant.
and  the  National Natural Science Foundation of China (No. 61304181).

\begin{thebibliography}{99}
\item Y. Yuan, X.T. Zhuang, Multifractal description of stock price index fluctuation using a quadratic fitting, Physica A 387 (2008) 511-518.
\item P. Mali, A. Mukhopadhyay, Multifractal characterization of gold market: A multifractal detrended fluctuation analysis, Physica A 413 (2014) 361-372.

\item R. Hasan, M.Salim M, Multifractal analysis of Asian markets during 2007-2008 financial crisis, Physica A 419 (2015) 746-761.

\item M.B. Ribeiro, A.Y. Miguelote, Fractals and the Distribution of Galaxies, Brazilian Journal of Physics (1998) 132-134.

\item J. Li, Y. Chen, Rescaled range (R/S) analysis on seismic activity parameters, ACTA SEISMOLOGICA SINICA,Vol.14, No.2, (2001) 148-155.

\item K.E. Lee, J.W. Lee, Multifractality of the KOSPI in Korean stock market, J. Korean Phys. Soc. 46 (2005) 726-729.

\item Z. Chen, P.C. Ivanov, K. Hu, H.E. Stanley, Effect of nonstationarities on detrended fluctuation analysis, Phys. Rev. E 65 (4) (2002) 041107.

\item  G. Du, X. Ning, Multifractal properties of Chinese stock market in Shanghai, Physica A 387 (1) (2008) 261-269.

\item  J. Jiang, K. Ma, X. Cai, Non-linear characteristics and long-range correlations in Asian stock markets, Physica A 378 (2007) 399-407.

\item  P. Pavón-Domínguez, S. Serrano, F.J. Jiménez-Hornero, J.E. Jiménez-Hornero, E. Gutiérrez de Ravé, A.B. Ariza-Villaverde, Multifractal detrended fluctuation analysis of sheep livestock prices in origin, Physica A 392 (2013) 4466-4476.

\item J.W. Kantelhardt, S.A. Zschiegner, E.Koscielny-Bunde, S. Havlin, A.Bunde, H.E.Stanley, Multifractal detrended fluctuation analysis of nonstationary time series, Physica A 316 (2002) 87-114.

\item  L. Zunino, B.M. Tabak, A. Figliola, D.G. Perez, M. Garavaglia, O.A. Rosso, A multifractal approach for stock market inefficiency, Physica A 387 (2008) 6558-6566.

\item P.Norouzzadeh, G.R. Jafari, Application of multifractal measures to Tehran price index, Physica A 356 (2005) 609-627.

\item K.Matia, Y.Ashkenazy, H. E.Stanley, Multifractal properties of price fluctuations of stock and commodities, Europhysics Letters, Vol.61, NO.3 (2003) 422-428.

\item P.Suarez-Garcia, D.Gomez-Ullate, Multifractality and long memory of a financial index, Physica A 394 (2014) 226-234.

\item A.Bashan, R.Bartsch, J.W.Kantelhardt, S.Havlin, Comparison of detrending methods for fluctuation analysis, Physica A 387 (2008) 5080-5090.

\item Y. Yuan, X.T. Zhuang, X. Jin, Measuring multifractality of stock price fluctuation using multifractal detrended fluctuation analysis, Physica A 388 (11) (2009) 2189-2197.

\item W.X. Zhou, The components of empirical multifractality in financial returns, Europhys. Lett. 88 (2009) 28004

\item H.G.Chen, D.Q.Zhou, Multifractal Spectrum Analysis of Crude Oil Futures Prices Volatility in NYMEX, Management and Service Science (2010).

\item F.Wang, G.P.Liao, J.H.Li, X.C.Li, T.J.Zhou, Multifractal detrended uctuation analysis for clustering structures of electricity price periods, Physica A 392 (2013) 5723-5734.

\item N.Aslanidis, S.Fountas, Is real GDP stationary? Evidence from a panel unit root test with cross-sectional dependence and historical data, Empir Econ, 46 (2014) 101-108.

\end {thebibliography}
\end{document}